\documentclass[%
 reprint,
 superscriptaddress,
 amsmath,amssymb,
 aps,
 floatfix,
]{revtex4-2}
\usepackage[utf8]{inputenc}
\usepackage{graphicx}
\usepackage{subcaption}
\usepackage{dcolumn}
\usepackage{bm}
\usepackage{booktabs}
\usepackage{xfrac}
\usepackage[normalem]{ulem}
\usepackage{caption}
\usepackage{amsmath}

\begin{document}

\preprint{APS/123-QED}

\title{Efficient Monte Carlo Event Generation for Neutrino-Nucleus Exclusive Cross Sections}

\author{Mathias El Baz}
\email{mathias.elbaz@unige.ch}
\affiliation{Département de Physique Nucléaire et Corpusculaire, Université de Genève, Geneva, Switzerland}
\author{Federico Sánchez}
\affiliation{Département de Physique Nucléaire et Corpusculaire, Université de Genève, Geneva, Switzerland}
\author{Natalie Jachowicz}
\affiliation{Department of Physics and Astronomy, Ghent University, Proeftuinstraat 86, 9000 Gent, Belgium}
\author{Kajetan Niewczas}
\affiliation{Department of Physics and Astronomy, Ghent University, Proeftuinstraat 86, 9000 Gent, Belgium}
\author{Ashish Kumar Jha}
\affiliation{Department of Physics and Astronomy, Ghent University, Proeftuinstraat 86, 9000 Gent, Belgium}
\author{Alexis Nikolakopoulos}
\affiliation{Fermi National Accelerator Laboratory, Batavia, Illinois, USA}
\affiliation{Department of Physics, University of Washington, Seattle, WA 98195, USA}


\date{\today}

\begin{abstract}
Modern neutrino-nucleus cross section computations need to incorporate sophisticated nuclear models to achieve greater predictive precision. However, the computational complexity of these advanced models often limits their practicality for experimental analyses. To address this challenge, we introduce a new Monte Carlo method utilizing normalizing flows to generate surrogate cross sections that closely approximate those of the original model while significantly reducing computational overhead. As a case study, we built a Monte Carlo event generator for the neutrino-nucleus cross section model developed by the Ghent group. This model employs a Hartree-Fock procedure to establish a quantum mechanical framework in which both the bound and scattering nucleon states are solutions to the mean-field nuclear potential. The surrogate cross sections generated by our method demonstrate excellent accuracy with a relative effective sample size of more than $98.4 \%$, providing a computationally efficient alternative to traditional Monte Carlo sampling methods for differential cross sections.
\end{abstract}

\maketitle

\section{\label{sec:level1}Introduction}
The study of neutrino interactions with nuclei is critical for reducing systematic uncertainties in modern neutrino oscillation experiments. Among the most important processes is Charged Current Quasi-Elastic (CCQE) scattering, where a neutrino interacts with a nucleon target inside a nucleus, emitting a charged lepton while ejecting one nucleon. This interaction forms the backbone of event reconstruction in neutrino oscillation experiments like T2K (Tokai to Kamioka experiment~\cite{T2K:2011qtm}) and the new Hyper-Kamiokande water Cherenkov detector~\cite{protocollaboration2018hyperkamiokandedesignreport}. However, the theoretical modeling of CCQE interactions is complex due to the need to account for the nuclear environment in which the nucleons are bound originally~\cite{NuSTEC:2017hzk}.

Moreover, these experiments often use neutrino beams that span a broad energy spectrum—from a few hundred MeV to several GeV—making Monte Carlo (MC) event generators indispensable. MC generators, such as NEUT~\cite{Hayato:2009zz}, GENIE~\cite{Andreopoulos:2009rq}, NuWro \cite{Golan:2012wx} and ACHILLES \cite{Isaacson_2023}, efficiently sample different interaction processes over many possible final states, enabling exhaustive modeling across this wide energy range. Such modeling is essential for extracting oscillation parameters and minimizing systematic uncertainties in neutrino physics.

A variety of interaction models have been implemented in MC generators to describe neutrino-nucleus interactions. However, most of these models are limited to providing only inclusive cross sections and therefore cannot reliably predict final-state hadron kinematics. As a consequence, the latter are often generated in an ad hoc and unrealistic manner~\cite{Dolan_2020, nikolakopoulos2023inclusivesemiinclusiveonenucleonknockout}.

The only approaches currently used in neutrino generators that systematically provide final-state nucleon kinematics are based on the factorization obtained in the plane-wave impulse approximation (PWIA). In this picture, one assumes no final-state interactions, so the process can be factorized into (i) the primary neutrino-nucleon interaction and (ii) a nuclear model describing the bound nucleons. However, these approaches lack the effects of the nuclear medium on the outgoing nucleon. 

In the simplest instance of PWIA, the nucleus is modeled as a Relativistic Fermi Gas (RFG)~\cite{PhysRevC.72.025501}, in which nucleons are treated as  fermions uniformly occupying momentum states up to the Fermi momentum  in a constant potential. This approach has been remarkably successful in describing general properties of intermediate-energy processes, especially in inclusive electron scattering. Yet, it neglects important nuclear features such as the shell structure and nucleon-nucleon correlations.

A more refined description of the nuclear medium still within the PWIA is provided by the Spectral Function (SF) model~\cite{BENHAR1994493}, which yields a probability distribution $ S(p,E) $ for finding a nucleon with momentum $ p $ and removal energy $ E $ based on $(e,e'p)$ data and theoretical inputs. Although the SF model captures the spread in binding energies and some nucleon-nucleus correlations, its essential reliance on the PWIA 
means that correlations essential to accurately predict the final-state hadron kinematics cannot be fully taken into account.

While these approximations were adequate when the focus was largely on lepton kinematics, the increased precision of upcoming accelerator-based long-baseline neutrino oscillation experiments will enable much more detailed measurements of outgoing hadron kinematics. To achieve the few-percent level of uncertainty on neutrino interactions required by these next-generation neutrino oscillation studies, cross section models must go beyond PWIA to include final-state interactions. 

To obtain exclusive differential cross sections, a theoretical framework that accurately describes the hadron system is required. One such approach is based on the distorted-wave impulse approximation (DWIA) in the mean field approximation ~\cite{RING1996193, PhysRevC.65.025501}. Mean-field-based models describe nucleons within an averaged nuclear potential generated by all other nucleons. This naturally leads to a shell structure, where each nucleon in the initial state is described by a bound wave function. Meanwhile, the distorted-wave treatment goes beyond the plane-wave description by also considering the final-state interactions of the outgoing nucleon: instead of emerging as a free particle, the outgoing nucleon propagates through the same mean field potential and undergoes distortions due to its interactions with the residual nucleus. This approach has recently gained renewed attention in the context of event generator development, with the first implementation of a relativistic DWIA-based CCQE model in NEUT~\cite{mckean2025implementationrelativisticdistortedwave}.

Despite their advantages, mean-field-based cross section calculations are computationally expensive compared to the factorized PWIA. The slow evaluation speed of these cross sections translates into slow sampling rates when traditional methods like accept-reject algorithms are used. To mitigate this, some current solutions \cite{Nikolakopoulos_2022, Nikolakopoulos_2024, Douqa_2024, Gonz_lez_Jim_nez_2022}  involve precomputing the hadron tensor components into heavy multidimensional grids, known as hadron tensor tables \cite{Dolan_2020, Dolan:2021rdd, schwehr2017genieimplementationificvalencia, Prasad:2024gnv}. This method allows to compute the cross section much faster, and make an accept-reject sampling possible. However, this approach has significant drawbacks: it lacks flexibility, requiring recalculation for intricate change in theoretical parameters, and seem infeasible for multi-nucleon knockout exclusive cross sections due to the exponential scaling of table size with dimensionality.

Recent advancements in Artificial Intelligence, particularly in Deep Learning, offer promising solutions to these challenges \cite{Heimel_2024, graczyk2024electronnucleuscrosssectionstransfer, Butter_2023, bonilla2025generativeadversarialneuralnetworks}. In this work, we propose an alternative sampling method based on normalizing flows (NF)~\cite{rezende2016variationalinferencenormalizingflows}. Normalizing flows transform a simple base distribution, such as a Gaussian, into a complex target distribution through a series of learnable diffeomorphisms. This enables efficient sampling and evaluation of the target distribution—in this case, the neutrino-nucleus cross section. In this work, we demonstrate the capacity of NF to build a Monte Carlo event generator for neutrino-induced cross sections. For our proof-of-principle we use the Hartree-Fock mean-field exclusive 1-particle-1-hole ($1p1h$) cross section formalism developed by the Ghent group~\cite{RYCKEBUSCH1994828, PhysRevC.51.2664, PhysRevC.65.025501}.   
These cross sections exhibit all the features needed to demonstrate that our approach can pave the way for faster and more flexible event generation in neutrino physics.

The structure of this article is as follows: we first review the theoretical framework behind CCQE interactions within the mean-field approach adopted by the Ghent group. Following this, we introduce the normalizing flows technique and outline how it is integrated into the Monte Carlo event generation process. Finally, we present results demonstrating the efficiency and accuracy of the generator.

\section{Neutrino-nucleus exclusive CCQE cross section in the Mean Field Framework}
\label{cross section in mf}

The aim of this study is to demonstrate the ability to generate events with precision and speed for single nucleon knockout distributed according to the Hartree-Fock calculations developed by the Ghent group.  While this model can predict both single and multiple nucleon knockout, providing a Monte Carlo event generator for all processes in their full complexity lies beyond the scope of this work. Future steps for handling the other processes are discussed in Section~\ref{discussion}. Here, we focus on $1p1h$ interactions due to a single-nucleon operator. The goal of this section is to introduce the relevant kinematic structure of the cross section.

\begin{figure}[h]
    \centering
    \includegraphics[width=1\linewidth]{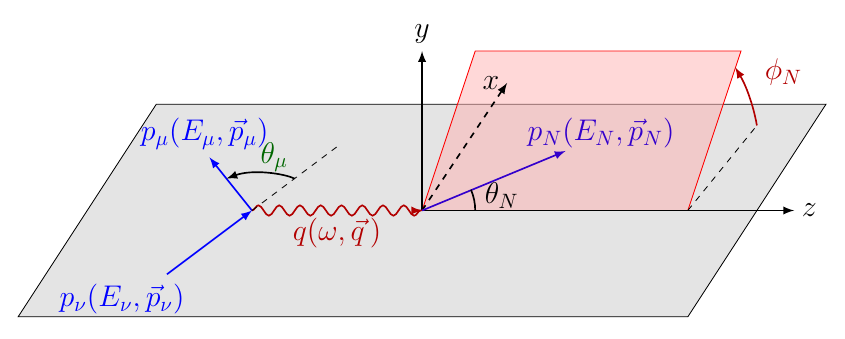}
    \caption{1p1h kinematics in the laboratory frame with the $z$ axis along the momentum transfer}
    \label{fig:1p1h}
\end{figure}

The process is the following:
\begin{equation}
\nu_\mu + {^A_Z X} \;\longrightarrow\; \mu^- \;+\; p \;+\; \bigl(^{A-1}_{Z-1}X\bigr)^*,
\end{equation}
where:
\begin{itemize}
    \item $\nu_\mu$ is the incoming muon neutrino,
    \item ${^A_Z X}$ represents the target nucleus with atomic number $Z$ and mass number $A$,
    \item $\mu^-$ is the outgoing charged lepton,
    \item $p$ is the ejected proton,
    \item $\bigl(^{A-1}_{Z-1}X\bigr)^*$ is the residual nucleus in an excited state due to the interaction.
\end{itemize}
The kinematics are displayed in Figure~\ref{fig:1p1h}, here the target nucleus is at rest, and the coordinate system is chosen so that the momentum transfer $\vec{q}$ defines the $z$-axis.

The interaction between the neutrino and a nucleon within the nucleus by one-boson exchange is described using the contraction of the lepton and hadron tensors, denoted as $ L_{\mu\nu} $ and $ W^{\mu\nu} $, respectively:
\begin{equation*}
\frac{d^6\sigma (E_\nu)}{d\omega \, d\Omega_\mu \, dE_N \, d\Omega_N} \propto L_{\mu\nu} W^{\mu\nu}.
\end{equation*}
where $\omega = E_\nu - E_\mu$ is the energy transfer, $E_N$ the outgoing nucleon energy, and $\Omega$ the solid angles for the muon and proton.
The lepton tensor is independent of the nuclear model, and is calculated in electroweak theory.  
As the cross section is not differential in neutrino energy, event generators typically treat $ E_\nu $ as an external input, modeling the conditional cross section for a fixed neutrino energy.

Clearly, the cross section is invariant with respect to rotations around the $ z $-axis. This allows samples to be uniformly rotated by a $ \phi $ angle in the range $ [0, 2\pi] $ without altering the physical results. Consequently, the variables that need to be sampled at fixed $E_\nu$ are $(\omega, \theta_\mu, E_N, \Omega_N)$. 

We make an additional simplification (without incurring any loss of information) to reduce the number of independent variables which is specific to the mean-field model. The hadron tensor $ W^{\mu\nu} $ is a bilinear product of the hadron currents with an energy conserving delta function :
\begin{equation}
W^{\mu\nu} = \overline{\sum_{i,f}} \langle i | \mathcal{J}^{\mu\dagger} | f \rangle \langle f | \mathcal{J}^\nu | i \rangle \delta(E_N + E_{A-1}^* - M_A - \omega),
\end{equation}
where $ |i\rangle $ and $ |f\rangle $ are the initial and final hadron states. The summation is understood to run over all relevant final states and average over the initial states. $\mathcal{J}^\nu$ is the nuclear current operator.

The nuclear states $ |i\rangle $ and $ |f\rangle $ are obtained as Slater determinants, composed of nucleon solutions to the Schrödinger equation with the mean-field potential $ V_{\text{MF}}^{(\alpha)} $:
\begin{equation}
\bigl[\hat{T}^{(\alpha)}(r) \;+\; V_{\mathrm{MF}}^{(\alpha)}(r)\bigr]\;\phi_\alpha(r)
\;=\;
\epsilon_\alpha\;\phi_\alpha(r)
\label{schrodinger mean field}
\end{equation}
with the kinetic energy operator being: 
\begin{equation*}
\hat{T}^{(\alpha)}(r)
\;=\;
-\frac{\hbar^2}{2}
\left[\frac{1}{r^2} \frac{d}{dr}
\bigl(\frac{r^2}{m^*(r)} \frac{d}{dr}\bigr)
-\frac{\ell_\alpha \bigl(\ell_\alpha + 1\bigr)}{ m^*(r) \, r^2}\right]
\end{equation*}
with $\alpha$ denoting the set of quantum numbers  identifying the single-nucleon state and $m^*(r)$ the effective mass (since the Hartree-Fock Hamiltonian is non-local)~\cite{PhysRevC.5.626}.
The mean-field potential is derived via a Hartree-Fock procedure~\cite{heyde2012nuclear}, starting from an effective Skyrme nucleon-nucleon potential~\cite{WAROQUIER1983298} and extending it to a nucleon-nucleus potential. 

The final state $ |f\rangle $ is generally expressed as a sum of orthogonal continuum solutions to the same equation with the correct asymptotic behavior. The complexity of nuclear modeling is embedded in the summation over all possible final states, which constitutes the primary computational bottleneck due to the necessity of computing and summing over a large number of distorted wave solutions of Equation \ref{schrodinger mean field}.

The ground state  $ |i\rangle $ is described as a Slater determinant of the $A$ lowest energy single-particle states of the mean field potential.
The states have quantum numbers $(\epsilon, n , l , j, m_j, i_z)$, i.e. respectively energy, principal quantum number, orbital and total angular momentum, projection of total angular momentum and isospin.
Due to spherical symmetry, the states within a shell characterized by quantum numbers 
$\alpha = (\epsilon_\alpha, n_\alpha, l_\alpha, j_\alpha, i_z)$ are energy-degenerate.
Since we are considering nucleon knockout to the continuum, the residual system can then only be left in states with excitation energy given by the single-particle energies of the shells $\epsilon_\alpha = M_N + E_{A-1}^* - M_A$.
Using this to integrate the delta function, and summing over $m_j$ the cross section decomposes into a sum of partial cross sections,
\begin{equation}
\label{eq:shells}
\frac{d^4\sigma (E_\nu)}{d\omega  \, d\theta_\mu \, d\Omega_N} = \sum_{\alpha} (2j_\alpha +1) \frac{d^4\sigma_{\alpha} (E_\nu)}{d\omega \, d\theta_\mu \, d\Omega_N},
\end{equation}
\label{sum partial}
where the sum runs over the shells with occupation $(2j_\alpha+1)$.
Each partial cross section produces a different spectrum of final-state nucleon energies determined by $\alpha$. 

We thus generate samples of the four independent continuous variables $(\theta_\mu, \omega, \theta_N, \phi_N)$ for each shell separately, with $E_\nu$ as an input. Among these variables, only the energy transfer $ \omega $ has a more complex marginal distribution, particularly at low values where nuclear effects play a significant role. This is due to the Hartree-Fock method preserving the shell structure of the nucleus  which in return impacts the cross section. In contrast, the remaining variables are angular and exhibit relatively smooth distributions over well-defined, bounded ranges. 

We lastly note here, that the dependence of the cross section on the azimuthal angle of the hadron plane, as shown in Fig.~\ref{fig:1p1h}, can always be decomposed into 5 simple functions~\cite{PhysRevD.90.013014, PhysRevD.98.073001}. 
For nucleon knockout the dependence on the nucleon's azimuthal angle can be written as
\begin{align}
\frac{d^4\sigma_\alpha (E_\nu)}{d\omega \, d\theta_\mu \, d\Omega_N} &= \frac{d^3\sigma_\alpha (E_\nu)}{d\omega \, d\theta_\mu \, d\cos\theta_N} \nonumber \\
&+B \cos\phi_N + C\cos 2\phi_N \nonumber \\
&+D \sin \phi_N + E \sin 2\phi_N,
\end{align}
in terms of the angle integrated cross section and the functions $B,C,D,E$ that only depend on $E_\nu, \omega,\theta_\mu, \theta_N$ and the shell.
This decomposition can be exploited to sample the $\phi_N$ dependence~\cite{Niewczas:2020fev}, and hence reduce the number of non-trivial independent variables.
For this work however, we sample the full kinematics directly from the partial cross sections of Eq.~(\ref{eq:shells}) .

\section{Event generator using normalizing flows}
\label{Event Generator using Normalizing Flows}

\subsection{Definition of normalizing flows}

Normalizing Flows (NF) are a class of machine learning models used to transform simple probability distributions into complex ones through a sequence of invertible and differentiable transformations of the probability space. They are particularly useful for efficiently sampling from high-dimensional and non-Gaussian distributions, making them highly suited for generating events from complex interaction models, such as neutrino-nucleus interactions. This contrasts with the accept-reject loop used in most neutrino event generators, which discards a large fraction of generated samples. 

The core idea behind NF is to start with a simple base distribution, typically a multidimensional Gaussian or uniform distribution, and apply a series of transformations that map this distribution into the more complex target distribution of interest.  Mathematically, a typical NF transformation consists of a sequence of $K$ diffeomorphisms of the probability space, each parameterized by a neural network, that progressively transform the base distribution $ p_Z(z) $ into the target distribution $ p_X(x) $. If $ z$ is a sample from the base distribution $ p_Z(z) $, the final output $ x = f_K \circ f_{K-1} \circ \cdots \circ f_1(z) $ represents a sample from the target distribution $ p_X(x) $.

To compute the likelihood $ p_X(x) $, we use the change of variables formula:
$$
p_X(x) = p_Z(z) \left| \det \frac{\partial f^{-1}}{\partial x} \right|,
$$
where $ z = f^{-1}(x) $ is the pre-image of $ x $ under the inverse transformation, and $ \det \frac{\partial f^{-1}}{\partial x} $ is the Jacobian determinant of the transformation. One can see the main advantage of NF for event generation: they allow to both sample and evaluate the probability of an event at the same time in an efficient and accurate way.

\subsection{Normalizing flows for event generation}

NF have been successfully applied to event generation in high-energy physics, particularly at the Large Hadron Collider (LHC). They have been used as a more efficient alternative to traditional importance sampling algorithms like VEGAS \cite{PETERLEPAGE1978192} to reduce the variance of an integral estimate. Gao et al.~\cite{Gao:2020zvv, Gao_2020} introduced an NF-based integrator called \texttt{iflow} to improve unweighting efficiency in Monte Carlo event generators, using Drell-Yan processes at the LHC as a case study. Bothmann et al.~\cite{Bothmann_2020} proposed a similar NF architecture applied to top-quark pair production and gluon scattering into three- and four-gluon final states. Stienen and Verheyen~\cite{10.21468/SciPostPhys.10.2.038} explored the use of autoregressive flows for efficient generation of particle collider events, performing experiments with leading-order top pair production events at an electron collider and next-to-leading-order top pair production events at the LHC. Further works, such as MadNIS developed by Heimel et al.~\cite{Heimel_2023, Heimel_2024} , use NF in a more advanced way \cite{Deutschmann:2024lml, Kofler:2024efb}.Collectively, these works demonstrate that NF significantly enhance unweighting efficiency in LHC event generation. Building upon these advancements, we aim to adapt and extend the use of NF to neutrino-nucleus interactions, providing an efficient event generator for complex neutrino-nucleus cross sections.

Normalizing flows rely on invertible transformations that allow for efficient application of the change-of-variable formula. Two types of architectures show the best performances: autoregressive flows and coupling layers. Both approaches exploit triangular (or block-triangular) Jacobians to ensure computational efficiency, with complexity scaling linearly with the dimensionality of the probability space.  However, the choice between autoregressive flows and coupling layer-based flows ultimately depends on the application.

 Autoregressive flows such as inverse autoregressive flow (IAF)~\cite{kingma2017improvingvariationalinferenceinverse} or masked autoregressive flow (MAF)~\cite{papamakarios2017masked} are D times slower to invert than to evaluate, where D is the dimension of the probability space.   On the other hand, flows based on coupling layers, such as NICE~\cite{dinh2015nicenonlinearindependentcomponents} or RealNVP~\cite{dinh2017densityestimationusingreal},  have an analytic one-pass inverse.  Sampling and evaluating the density at the same time requires to evaluate both the forward and inverse transformations. Therefore, coupling layers allow to both evaluate and sample a density fast. However, coupling layer-based flows are generally less expressive than autoregressive flows. Therefore, we chose to use autoregressive flows due to their higher expressiveness. A more recent comparative study between coupling and autoregressive flows by Coccaro et al.~\cite{Coccaro_2024} indicates that autoregressive flows stand out both in terms of accuracy and training speed. We will later also demonstrate that autoregressive flows offer sufficient evaluation and sampling speeds for the dimensionality of typical CCQE events.   
 
 In this study, we employ MAF, which use masked feed-forward networks to parameterize the NF transformations while enforcing their autoregressive property. The masking mechanism involves applying a binary mask to the weight matrices of the network, effectively setting specific connections to zero. This ensures that the output for a given dimension depends only on its predecessors in the ordering, preserving the conditional dependency property required for autoregressive modeling.  A permutation of the dimensions after each autoregressive transformation ensures that all dimensions are treated equally.

 \subsection{Circular Rational-Quadratic Neural Spline Flows}
 \label{Circular RQ-NSF}

In addition to using autoregressive flows to implement our NF,  we need to define the parametrization of the flow transformation.  In this study, we use Rational Quadratic Neural Spline Flows (RQ-NSF)  introduced by Durkan et al.~\cite{durkan2019neural}, which effectively models complex, non-Gaussian target distributions by utilizing piecewise rational quadratic transformations, as parametrized by Gregory and Delbourgo~\cite{delbourgo1983c}. RQ-NSF are very expressive due to their infinite Taylor-series expansion while being defined by a small number of parameters.  RQ-NSF are widely used in recent applications of NF for their state-of-the-art expressiveness. They have already been applied in the T2K collaboration for tasks such as modeling simple cross sections~\cite{PhysRevD.102.013003} and posterior density estimation for near-detector fits~\cite{PhysRevD.109.032008}. 

However, standard RQ-NSF, defined on Euclidean probability spaces, are suboptimal for modeling periodic variables, such as angular distributions. These distributions often exhibit sharp cut at boundaries when projected onto flat spaces, despite maintaining periodic boundary conditions. To address this, it is essential to model angles within their natural phase spaces—circles for angles and spheres for solid angles. In this work, we adopt the extension of RQ-NSF to non-Euclidean probability spaces, such as tori and spheres, as developed by Rezende et al.~\cite{rezende2020normalizingflowstorispheres}. By modifying Rational Quadratic Spline Flows to ensure periodicity in certain dimensions, we can appropriately model the periodic distributions through differentiable transformations of the probability space. In our specific $1p1h$ cross section model, the first two kinematics, $\theta_\mu$ and $\omega$,  can be represented on a cylindrical manifold due to the periodicity of $\theta_\mu$. The remaining two variables, $\theta_N$  and $\phi_N$ , are naturally represented on a sphere. Therefore, the natural phase space $\mathcal{M}$ where the cross section is defined is a cylinder times a sphere. 

\subsection{Energy-dependent flows}

In neutrino–nucleus interactions, the differential cross section depends on both the kinematic variables and the incoming neutrino energy $ E_\nu $. Neutrino event generators, such as NEUT~\cite{Hayato:2009zz}, typically take $ E_\nu $ as an input parameter and generate samples weighted by the conditional cross section given $ E_\nu $, rather than modeling $ E_\nu $ as an extra dimension. Consequently, we condition the NF directly on $ E_\nu $ to mimic the functioning. 

We model a family of probability distributions continuously parameterized by $ E_\nu $. This approach ensures that the NF captures the underlying dependence of the cross section on energy without explicitly including $ E_\nu $ as part of the flow's dimensional space. Additionally, this eliminates the need to retrain the NF for different values of $ E_\nu $.

For the specific case of $1p1h$ interactions, the goal is to model the conditional probability 
$$
p(\theta_\mu, \, \omega, \,\theta_N, \,\phi_N \mid E_\nu, \, \alpha),
$$
over a wide range of $ E_\nu $. During the transformation, the flow modifies the kinematic variables $ (\theta_\mu, \, \omega, \, \theta_N, \, \phi_N) $ in a way that captures the energy dependence implicitly. Here, we model each initial shell $\alpha$ separately.

The conditional probability is computed by normalizing the full 4D differential cross section for a given shell $\alpha$ by the total cross section:
$$
p(\theta_\mu, \, \omega, \, \theta_N,  \,\phi_N \mid E_\nu \, , \alpha) = \frac{\frac{d^4\sigma_\alpha(E_\nu)}{ \, d\omega \, d\theta_\mu \, d\theta_N \, d\phi_N}}{\sigma_\alpha(E_\nu)}.
$$

To model $ \sigma_\alpha(E_\nu) $, a standard approach is to integrate the cross section over the other variables and fit the resulting function using a polynomial approximation:
$$
\sigma_\alpha(E_\nu) \approx P_\alpha(E_\nu).
$$

\subsection{Exhaustive training of the flow-based event generator}

To train NF, we typically optimize the model by minimizing the Kullback-Leibler Divergence (KL-D) between the true distribution $ p $ and the learned distribution $ q_\theta $ (where $\theta$ refers to the neural net parametrization of the surrogate distribution). The KL-D measures the difference between two probability distributions, but it is inherently asymmetric, which means it leads to different behaviors depending on whether we minimize the forward or reverse KL-D. Minimizing the forward KL-D, expressed as
$$
D_{KL}(p \| q_\theta) = \mathbb{E}_{x \sim p} \left[ \log \frac{p(x)}{q_\theta(x)} \right],
$$
can lead to mean-seeking behavior, where the learned distribution $ q_\theta $ covers the regions where the true distribution $ p $ is non-zero but may overestimate the distribution tails. Conversely, minimizing the reverse KL-D, expressed as
$$
D_{KL}(q_\theta \| p) = \mathbb{E}_{x \sim q_\theta} \left[ \log \frac{q_\theta(x)}{p(x)} \right],
$$
can result in mode-seeking behavior, where $ q_\theta $ focuses on some peaks in the probability but may ignore other regions or even other peaks.

To balance these tendencies, we use the symmetric KL-D, which is the average of these two:
$$
\text{D}_{\text{s}}(p , q_\theta) = \frac{1}{2} \left( D_{KL}(p \| q_\theta) + D_{KL}(q_\theta \| q) \right).
$$

However, direct computation of the KL-D is impractical for our use case, as it would involve sampling from the true distribution $ p $, which is computationally and time expensive. Furthermore, sampling from the learned distribution $ q_\theta $ to compute the loss function introduces instability. This instability arises because, in the early training stages, the learned distribution may deviate significantly from the target, leading to unreliable loss estimates.

A significant improvement in our training process was achieved through the application of importance sampling to compute the loss function~\cite{müller2019neuralimportancesampling}. Importance sampling employs a proposal distribution that is simple to sample from yet effectively spans the target distribution $p$ . In our case, we used a uniform distribution $u$ defined over the manifold $\mathcal{M}$, as defined in Section \ref{Circular RQ-NSF}, and sampled energy values uniformly within the chosen training range. This enabled us to compute a Monte Carlo estimate of the symmetric KL-D:
$$
\text{D}_{\text{s}}(p , q_\theta) \approx \frac{1}{Z}\sum_{i,j}\left[ \left( p(x_i|E_j) - q_\theta(x_i|E_j) \right) \log \frac{p(x_i|E_j)}{q_\theta(x_i|E_j)} \right],
$$
where $Z$ is a normalization constant, and $x_i$ and $E_j$ are sampled uniformly from $\mathcal{M}$ and the training energy range, respectively. 
This approach ensures robust and stable training by exhaustively covering the manifold, preventing the learned distribution $q_\theta$ from collapsing into narrow regions or missing areas. 

To contextualize, a related approach was proposed by Pina-Otey et al.~\cite{PhysRevD.102.013003} in their Exhaustive Neural Importance Sampling (ENIS) framework. They employed a two-stage training process: an initial "warm-up" phase, during which the training points were uniformly sampled for 20\% of the total training time, followed by direct sampling from the learned distribution to compute the loss function. The first stage allowed exhaustive exploration of the phase space, while the second stage accelerated convergence by focusing on regions where the model had already predicted probability density. 
\\
Although the two-stage strategy is appealing, our results indicate that maintaining uniform sampling throughout the training process is not only simpler but also sufficiently fast for the dimensionalities involved in the exclusive $1p1h$ processes. Furthermore, in scenarios where cross section calculations are computationally expensive, this approach becomes impractical, as it would require evaluating the cross section repeatedly during training for new samples of the predicted probability density. Exhaustivity remains a critical factor in cross section modeling, where the primary goal is to develop the most accurate and comprehensive surrogate for the cross section, rather than merely optimizing training time. However, for higher-dimensional processes like two-particle-two-hole ($2p2h$) interactions, uniformly sampling the kinematics may become infeasible. In such cases, an alternative approach could involve training the model using samples drawn from an already trained model for the semi-exclusive cross section, where the kinematics of the sub-leading nucleon are marginalized. This semi-exclusive distribution could then serve as a proposal distribution for training a normalizing flow to model the fully exclusive cross section.

\section{Performances of the flow-based event generator}

\subsection{Detail of implementation and training}
\label{detail of implementation}

We trained a flow-based Monte Carlo event generator for the CCQE $1p1h$ interaction between a neutrino and a carbon nucleus ($^{12}\text{C}$). The carbon nucleus consists of two shells, $1s_{\sfrac{1}{2}}$ and $1p_{\sfrac{3}{2}}$, with respective separation energies of $36.20$ MeV and $18.72$ MeV. To enhance the flexibility of our event generation, we chose to model the interactions for each shell separately. This approach serves theoretical purposes, enabling the study of interactions with individual shells. A similar work can be applied to any nuclei, which would involve modeling more or fewer shells.

Additionally, we trained four separate models per shell, each corresponding to overlapping neutrino energy ranges in MeV: $[180, 350]$, $[300, 500]$, $[450, 750]$, and $[700, 1050]$. Consequently, the event generator covers neutrino energies from $180$ MeV to $1050$ MeV. The overlapping regions between these energy ranges ensure smooth transitions between models. In these overlap regions, we blend the two overlapping models by sampling from and evaluating the second model with a probability that increases linearly from $0$ at the start of the overlap range to $1$ at the end. $^{12}\text{C}$ having $2$ shells, the total number of models is therefore $8$.

Concerning the implementation of a single model, we used an implementation of Autoregressive RQ-NSF based on the \texttt{nflows} Pytorch implementation of Durkan et al. ~\cite{nflowsgit}. Some modifications were made to accommodate the computation of the Symmetric Kullback-Leibler divergence using importance sampling. 

\begin{figure*}[ht!]
\centering
    \subcaptionbox{True cross section }
    [.49\linewidth]{\includegraphics[width = 0.99\linewidth, trim=0cm 0cm 0cm 0cm, clip]{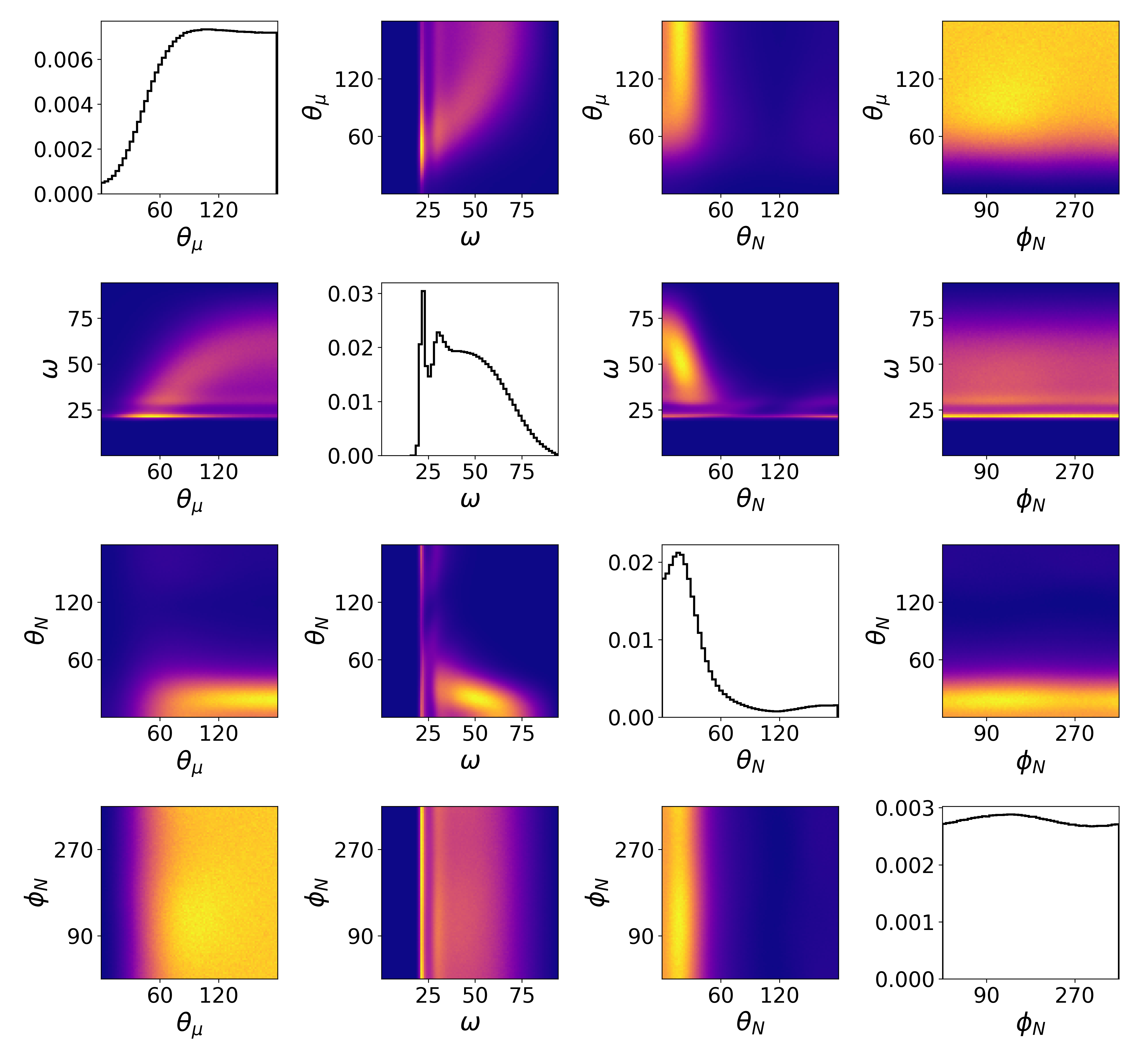}}
    \subcaptionbox{Surrogate cross section}
    [.49\linewidth]{\includegraphics[width = 0.99\linewidth, trim=0cm 0cm 0cm 0cm, clip]{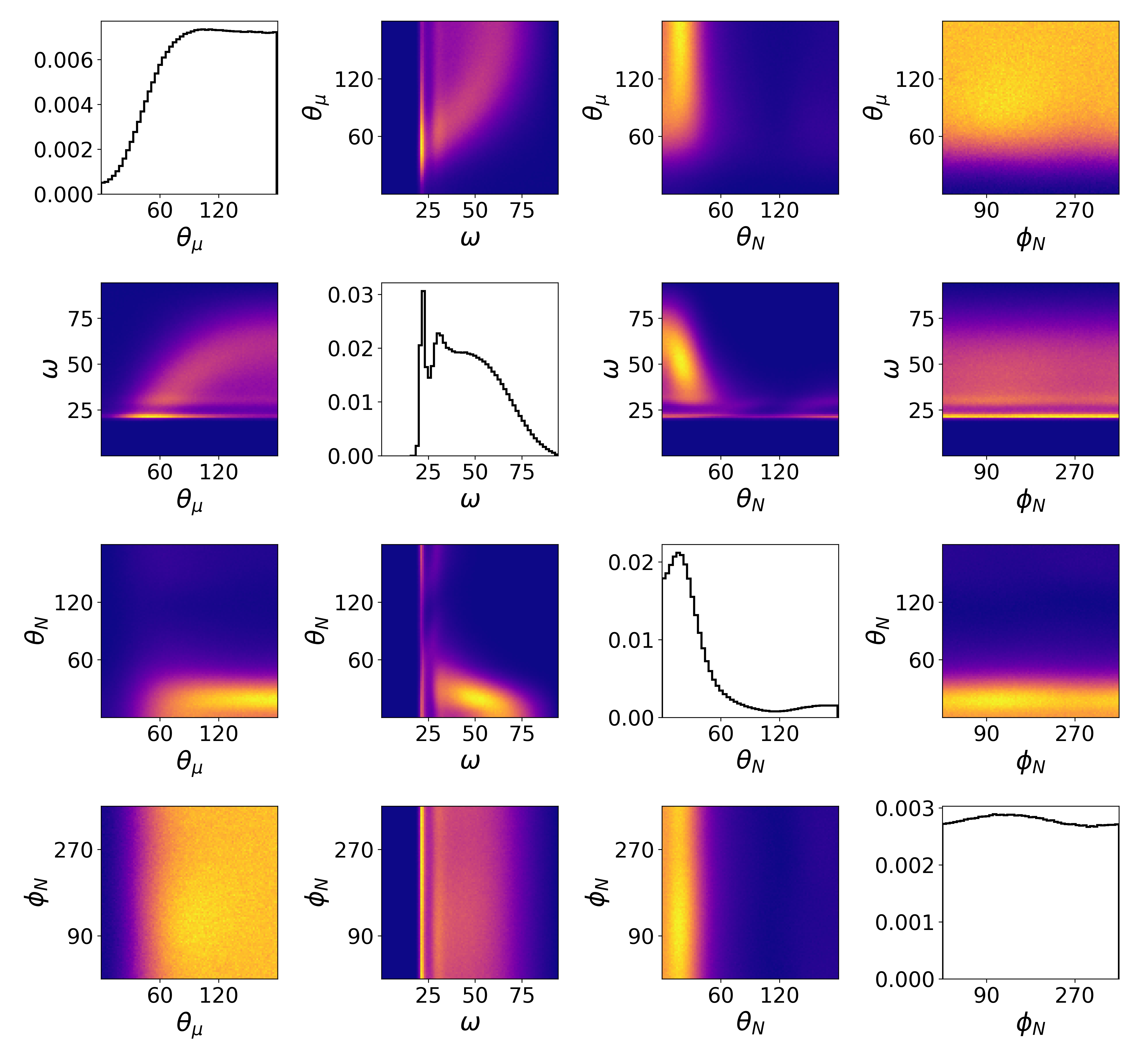}}
    \caption{100 million samples with weights from the true cross section (a) and 10 million weighted samples from the predicted surrogate cross section (b) for a neutrino energy of $E_\nu = 
 200 $ MeV and the $1p_{\frac{3}{2}}$ shell. The diagonal plots represent the marginal distribution of the 4 kinematic variables $(\theta_\mu, \omega, \theta_N, \phi_N)$. The off-diagonal plots correspond to the 2D histograms of 2 different kinematic variables.}
    \label{fig: 200 MeV, 1p}
\end{figure*}

We trained three types of models, each differing in size: a small model (\textbf{S}), a medium model (\textbf{M}), and a large model (\textbf{L}). All three models utilize $9$ bins in their spline transformations. Each RQ-NSF is parameterized by a masked autoregressive network composed of three hidden layers, with a residual connection from the first to the third layer. Model S includes $256$ hidden nodes per hidden layer and $10$ RQ-NSFs, Model M uses $512$ hidden nodes and $10$ RQ-NSFs, and Model L employs $512$ hidden nodes and $25$ RQ-NSFs. The performance of these three models is evaluated in Table \ref{tab:model_comparison}.   The base distribution is a four-dimensional probability density with independent dimensions. The three angular dimensions are modeled as uniform distributions, while the energy transfer dimension follows a normal distribution. 

We used a relatively low learning rate of $0.0001$ with a Cosine annealing scheduler with a minimum learning rate of $5 \times 10^{-7}$. We trained each model for $100000$ epochs. Each epoch computes the Symmetric KL-D on a batch of $2^{14}$ samples with associated true cross section. A total of $10$ million samples were used to train each model, with each sample appearing multiple times throughout the training. These samples and their cross sections were pre-generated prior to training—a potentially time-intensive step depending on the cross section computation speed, but required only once. The $\theta_\mu$ and $\theta_N$ distributions are unfolded to $[-\pi,\pi]$ to ensure $2\pi$-periodicity. The training time for each model (on an NVIDIA RTX-4090 GPU) are given in Table \ref{tab:model_comparison}.

\subsection{Comparison of the true cross section and its surrogate projecting in 1D and 2D spaces}

In Monte Carlo event generators widely used in neutrino physics, such as NEUT, users can freely select any distribution of neutrino energy. As a result, our model must accurately represent all possible combinations of energy and, in our case, the interacting nucleon's shell. While it would be impractical to visually compare the true and surrogate cross sections for every such combination, Sections \ref{Comparison Density Estimation} and \ref{Comparison Datasets} provide a more quantitative and exhaustive evaluation of the surrogate cross section's performance. This section offers a more visual and qualitative evaluation for a specific energy and shell combination.
 
Figure \ref{fig: 200 MeV, 1p} illustrates the comparison between the true cross section and its surrogate modeled using normalizing flows for a fixed neutrino energy of $E_\nu = 200$ MeV, with the initial bound nucleon in the $1p_{\frac{3}{2}}$ shell. This particular configuration was chosen because nuclear effects have a significant impact on its cross section. At low energy transfer, the cross section is especially sensitive to shell structure, a key feature captured by the Hartree-Fock mean-field approach.  The angular distribution of the outgoing nucleon relative to the transferred momentum, $ \theta_N $, is strongly influenced by the properties of the initial bound state. Nevertheless, the NF seem to model accurately this distribution from the shapes of the four marginal distributions to the correlations between two kinematic variables. 

\subsection{Comparison of the true cross section and its surrogate using density estimation}
\label{Comparison Density Estimation}

\subsubsection{Point-by-point density comparison}

To compare two multidimensional distributions, one often simplifies the problem by projecting the densities onto one or two dimensions at a time and/or by overlapping their binned histograms. These approaches, while visual, come with inherent limitations. By reducing the dimensionality, dependencies that might exist in higher-dimensional spaces are inevitably smoothed out. Similarly, binning locally averages out the densities and therefore can obscure fine-scale variations, introducing artificial agreement in regions where the true distributions may differ subtly. 

Therefore, one needs a more holistic approach to comparing multidimensional distributions. In this work, we evaluate directly the weight which is defined as the ratio between the true distribution $p$ and its normalizing flows surrogate $q_\theta$ (where $\theta$ refers to the neural net parametrization of the surrogate distribution):

$$w (\theta_\mu, \omega, \theta_N, \phi_N, E_\nu, \alpha) = \frac{p(\theta_\mu, \omega, \theta_N, \phi_N \mid E_\nu, \alpha)}{q_\theta(\theta_\mu, \omega, \theta_N, \phi_N \mid E_\nu, \alpha)} $$

\noindent where $\alpha$ refers to the bound nucleon shell. This is the error one expects in a sample if the NF surrogate was used in its simplest form, and also the weight one would want to apply to an event generator sample to correct the imperfect surrogate to be closer to the true distribution. 

\begin{figure}[h!]
    \centering
    \includegraphics[width=0.99\linewidth]{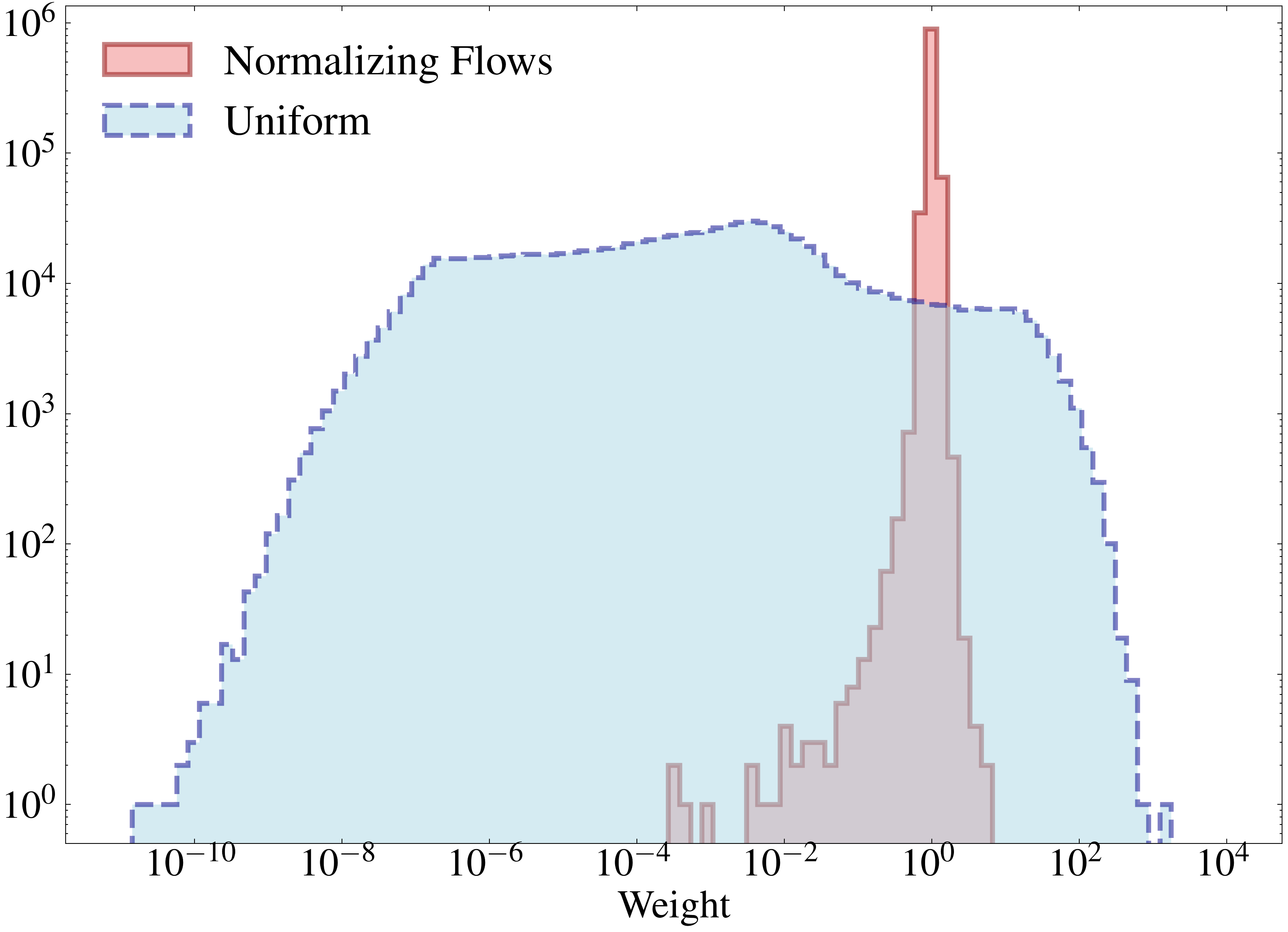}
    \caption{Log-scale histogram of $1$ million weights for both shells of $^{12}\text{C}$ under a flat neutrino energy flux from $200$ to $1000$ MeV, generated using the surrogate cross section (red line) and a uniform distribution (blue dashed line) defined on the manifold $\mathcal{M}$ .}
    \label{fig:weights}  
\end{figure}

The histogram of weights for the \textbf{L} model along with the histogram of weights for a uniform distribution defined on the manifold $\mathcal{M}$ are shown in Figure \ref{fig:weights}. In the case of the surrogate cross section, the weights are tightly concentrated around $1$, with a measured standard deviation of $10.7\%$ and a $99.99^\text{th}$ percentile value of $1.90$. This represents a significant advantage when compared to a uniform distribution, where the weights span $13$ orders of magnitude, leading to a large proportion of samples contributing minimally or not at all.

From these weights, we derive the Relative Effective Sample Size (RESS), which provides a quantitative measure of the quality of the surrogate distribution $q_\theta$ in approximating $p$. The RESS quantifies the proportion of effectively independent samples represented by the surrogate distribution when weighted by the importance weights. It is defined as:
$$
\text{RESS} = \frac{1}{N} \frac{\left(\sum_{i=1}^N w_i\right)^2}{\sum_{i=1}^N w_i^2},
$$
where $w_i$ are the weights for the samples $i = 1, \dots, N$.

A RESS close to $1$ indicates that the weights $w_i$ are similar in magnitude, reflecting that the surrogate distribution $q_\theta$ closely approximates the true distribution $p$. In contrast, a RESS close to $0$ suggests that the weights are spread out, implying a significant mismatch between $q_\theta$ and $p$. The performance of the models is evaluated using $1$ million samples drawn from $q_\theta$ across the full neutrino energy range, $[200, 1000]$ MeV, and for both shells combined. The RESS values for the three model sizes are presented in Table \ref{tab:model_comparison}. These values are computed for two of neutrino fluxes: one with a flat distribution and another based on the truncated T2K near-detector flux~\cite{T2kFlux}.

\subsubsection{Density comparison with finite resolution}
\label{density finite resolution}
Given that the RESS value obtained is slightly below $1$, it is necessary to assess the potential impact of the observed weight variation. A central question is whether the largest deviations are concentrated in specific regions of the phase space, which could introduce systematic biases in the generated sample exceeding current experimental uncertainties. Alternatively, these variations may stem from localized fluctuations which, when integrated over sufficiently small kinematic bins, effectively average out and thus remain negligible for experimental observations.

To clarify this point, the true and surrogate cross sections are both evaluated on a grid for a monoenergetic neutrino beam at $650$~MeV, summing both shell contributions. The grid spans different muon scattering angles with a step of $10^\circ$  and outgoing proton angles, specifically $\theta_N \in {1^\circ, 5^\circ, 10^\circ, 20^\circ, 30^\circ, 40^\circ, 50^\circ, 60^\circ}$ with a fixed nucleon azimuthal angle of $180^\circ$. The energy transfer $\omega$ is sampled with a step of $2.5$~MeV across its defined range. Importantly, the cross section values are averaged within localized four-dimensional kinematic bins centered at each point $(\omega, \theta_\mu, \theta_N, \phi_N)$. The resolution of each bin is $1$ MeV in energy transfer and $1^\circ$ in angular coordinates, which is significantly finer than what current experiments can resolve.  Every bin includes $500$ uniformly distributed samples, ensuring negligible statistical fluctuations in the averaged bin value.

\begin{figure}
\includegraphics[width=0.99\linewidth, trim=0cm 0cm 0cm 0cm, clip]{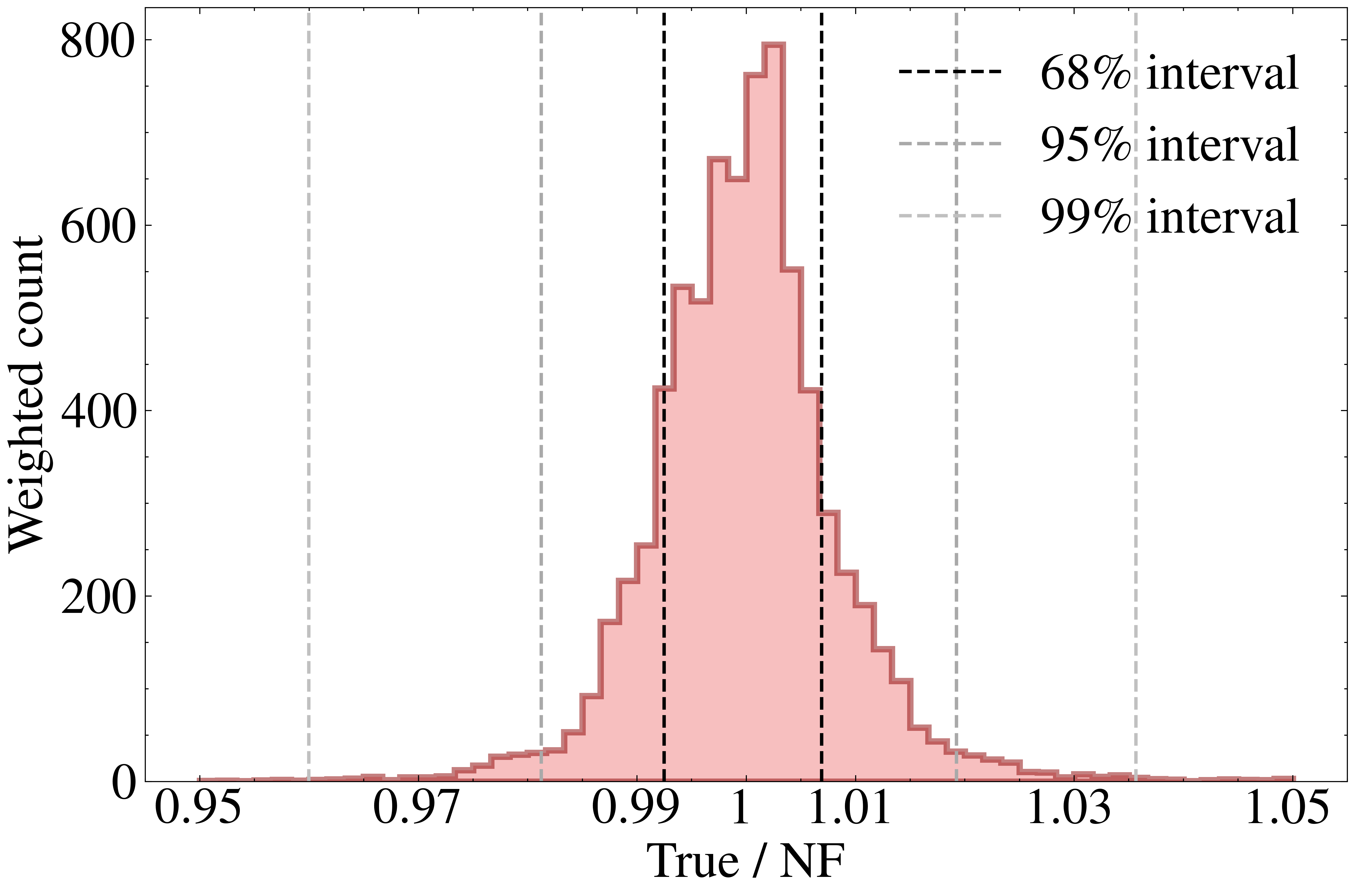}
\caption{
Histogram of the ratio between true and surrogate cross sections weighted by the true cross section evaluated at a neutrino energy of $650$ MeV across a fine kinematic grid and after local averaging. The dashed lines represent the confidence intervals: $68\%$ CI = ($-0.75\%$, $+0.69\%$), $95\%$ CI =  ($-1.88\%$, $+1.92\%$), and $99\%$ CI =  ($-4.00\%$, $+3.57\%$). 
}

\label{fig : hist avr}
\end{figure}

Figure \ref{fig : hist avr} presents the histogram of weights after averaging, of the given sampled points weighted by the true cross section values. The dashed lines represent the $68\%$, $95\%$, and $99\%$ confidence intervals.  Appendix~\ref{Profile appendix} further illustrates the impact of these cross section deviations on the distribution of proton kinetic energy $T_N$ at fixed kinematics, both with and without local averaging.

To quantify the residual discrepancies between the surrogate and true cross sections after local binning, we introduce the cross section-weighted root mean squared error (RMS$_p$), defined as:
\[
\text{RMS}_p = \sqrt{\frac{\sum_{\mathbf{x}_i} p(\mathbf{x}_i) \, \Delta^2(\mathbf{x}_i)}{\sum_{\mathbf{x}_i} p(\mathbf{x}_i) \, }},
\]
where:
\begin{itemize}
    \item $\mathbf{x}_i$ denotes a point in the kinematic space,
    \item $\Delta(\mathbf{x}_i)$ is the relative difference between the surrogate and true cross section at point $\mathbf{x}_i$ after averaging,
    \item $p(\mathbf{x}_i)$ is the true cross section at $\mathbf{x}_i$ after averaging.
\end{itemize}

We obtain an $\text{RMS}_p$ of $1.4\%$. This demonstrates that the local averaging procedure drastically reduces the modeling error compared to the initial spread of weights of $10.7\%$. The bias introduced by Normalizing Flows is therefore well below our experimental uncertainties. Furthermore, in realistic experimental conditions, where measurements involve a convolution over a broad neutrino energy spectrum, such residual differences would be further smeared out, rendering them more negligible in practice.

\subsection{Comparison between the true cross section and its surrogate through their respective datasets}
\label{Comparison Datasets}

In neutrino experiments such as T2K, we cannot achieve exact estimates of the kinematics we measure or infer. As a result, evaluating a model based on a direct point-by-point comparison of its distribution with the true one, as discussed in Section \ref{density finite resolution}, may be overly conservative. Instead, the goal in this work is to construct an event generator capable of populating the kinematic phase space in a manner consistent with a non-biased accept-reject method. This must be achieved with fixed precision while accounting for the full dimensionality of the phase space.

Therefore, we compare four-dimensional binned histograms derived from:
\begin{enumerate}
    \item Samples generated by the \emph{surrogate cross section}, and 
    \item Samples generated by an \emph{accept-reject algorithm} with a uniform proposal distribution defined on the manifold $\mathcal{M}$.
\end{enumerate}

We assess how closely these histograms align with a high-statistics, non-biased \emph{reference histogram} by computing the Multinomial Negative Log-Likelihood (MNLL):
\begin{equation*}
\text{MNLL}(H^{\text{MC}}) = \frac{1}{N} \sum_{\text{bin}} 
\Bigl[ \log{(H^{\text{MC}}_\text{bin}!)} - H^{\text{MC}}_\text{bin} \log{\bigl(\frac{H^{\text{ref}}_\text{bin}}{N}\bigr)}\Bigr],
\end{equation*}
where the summation runs over all bins in the histogram and:
\begin{itemize}
    \item $ H^{\text{MC}} $ is the histogram from the Monte Carlo dataset (either surrogate or accept-reject),
    \item $ H^{\text{ref}} $ is the high-statistics, non-biased reference histogram rescaled to match the number of samples in $ H^{\text{MC}} $,
    \item $ N = \sum_{\text{bin}} H^{\text{MC}}_\text{bin} $ is the total number of observed samples.
\end{itemize}

To obtain statistical distributions of the MNLL for both the surrogate model and the accept-reject method, we repeat the procedure $100$ times per method. Figure~\ref{fig:NLL} illustrates how the MNLL evolves with neutrino energy for both approaches. In this context, the unbiased accept-reject MNLL serves as the optimal performance benchmark for the surrogate cross section.

\begin{figure}[h!]
    \centering

    \begin{subfigure}[b]{0.99\linewidth}
        \centering
        \includegraphics[width=\linewidth]{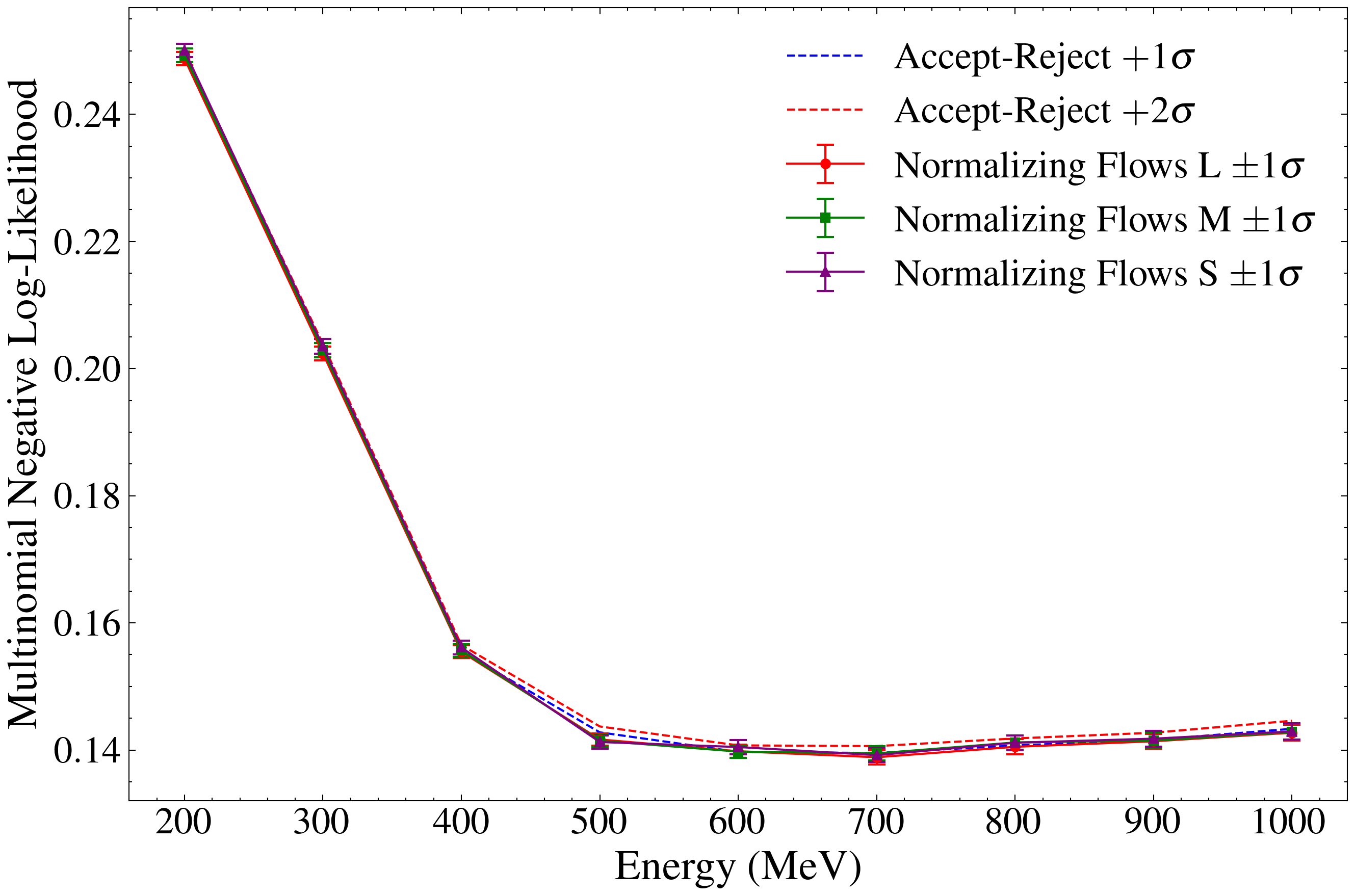}
        \caption{}
        \label{fig:NLL }
    \end{subfigure}
    \hfill
    \begin{subfigure}[b]{0.99\linewidth}
        \centering
        \includegraphics[width=\linewidth]{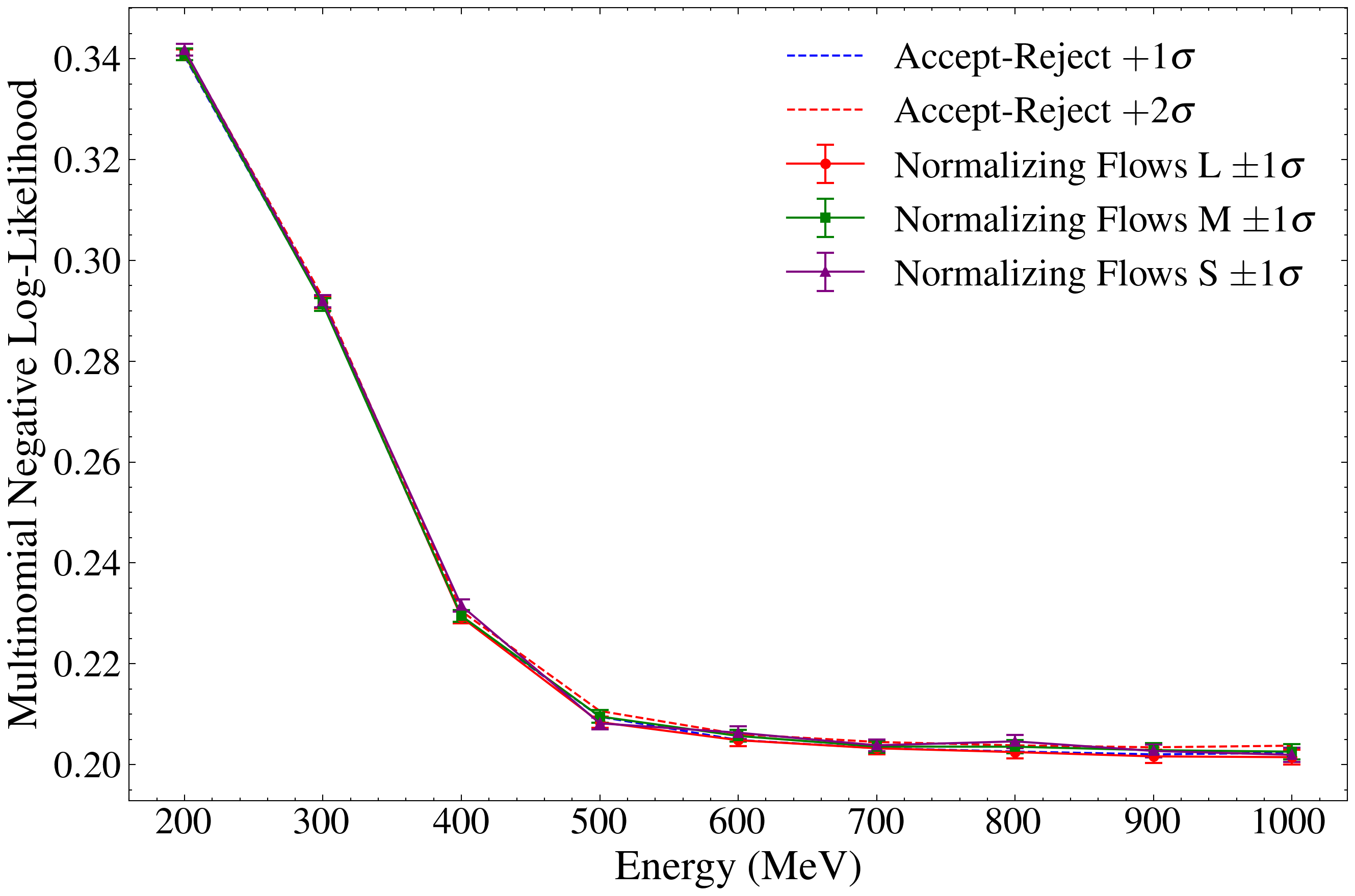}
        \caption{}
        \label{fig:NLL 1p}
    \end{subfigure}
    \caption{Evolution of the Multinomial Negative Log-Likelihood with the neutrino energy for the $1s_{\sfrac{1}{2}}$ ((a)) and $1p_{\sfrac{3}{2}}$ ((b)) shell. Each plot shows both the MNLL for the for the three normalizing flows model size. The $+1 \sigma$ and  $+2 \sigma$ upper bounds of the MNLL using the Accept-Reject datasets are given for comparison.}
    \label{fig:NLL}
\end{figure}

This comparison is carried out at fixed neutrino energies of 200, 300, 400, 500, 600, 700, 800, 900 and 1000 MeV, considering the $1s_{\sfrac{1}{2}}$ and $1p_{\sfrac{3}{2}}$ shells separately. For each energy and shell, $\theta_\mu$ and $\theta_N$ are binned with a width of $10^\circ$, while $\phi_N$ has a bin width of $30^\circ$. For the energy transfer, the bin width is $10\,\text{MeV}$ from the separation energy up to $100\,\text{MeV}$, and then $50\,\text{MeV}$ until reaching the maximum energy transfer of $E_\nu - m_\mu$, where $E_\nu$ is the neutrino energy and $m_\mu = 105.66\,\text{MeV}$ is the muon rest mass. We choose a finer binning at lower energy transfer to capture the complex structure inherited from the shell modeling. Each histogram $H^{\mathrm{MC}}$ uses $100{,}000$ samples, while the reference histogram is built from 10 million weighted samples.

To provide a measure of how the surrogate’s MNLL values differ from those of the accept-reject method in average, we compute the  RMS of the Z-Score ($\text{RMS}_z$) across all energies for a given shell:

\begin{equation*}
\begin{split}
\text{RMS}_z = 
\sqrt{\frac{1}{N_\text{e}} 
\sum_{i=1}^{N_\text{e}} 
\left( 
\frac{\overline{\text{MNLL}(H^{\text{NF}}_{E_\nu^{(i)}})} - 
\overline{\text{MNLL}(H^{\text{AR}}_{E_\nu^{(i)}})}}
{\sigma^\text{AR}_{E_\nu^{(i)}}} 
\right)^2 }
.
\end{split}
\end{equation*}
where 
$\overline{\text{MNLL}(H^{\text{NF}}_{E_\nu^{(i)}})}$ is the mean MNLL for normalizing flow  datasets at energy $E_\nu^{(i)}$, 
$\overline{\text{MNLL}(H^{\text{AR}}_{E_\nu^{(i)}})}$ is the mean MNLL for accept-reject  datasets at the same energy, 
and $\sigma^\text{AR}_{E_\nu^{(i)}}$ is the standard deviation of the accept-reject MNLL distribution at energy $E_\nu^{(i)}$. The $\text{RMS}_z$ for the three model sizes are given in Table \ref{tab:model_comparison}.  The MNLL distributions for the surrogate and accept-reject histograms are statistically compatible within their $\text{RMS}_z=1$ range for the $L$ model and within $\text{RMS}_z =  2$ for the $M$ and $S$ models. In the case of the model size \textbf{L}, we find an average $\text{RMS}_z$ of $0.75\,\sigma$ for the $1s_{\sfrac{1}{2}}$ shell  and $0.88\,\sigma$ for the $1p_{\sfrac{3}{2}}$ shell.  In other words, two datasets of $100,000$ samples respectively generated by accept-reject and by the normalizing flow surrogate are practically equivalent given the binning that we chose. Overall, these results demonstrate that the normalizing flow surrogate reproduces the cross section with high fidelity in \emph{all} tested combinations of shell and energy and in the full kinematic space, while significantly reducing computational overhead compared to the accept-reject method.

\subsection{Overall performance}

The performances for the three model sizes are summarized in Table \ref{tab:model_comparison}. While accuracy improves slightly with increasing model size, this comes at the expense of reduced computational efficiency and larger model sizes. The sampling speeds are here given for a single neutrino energy and shell combination. As noted earlier, we prioritized modeling accuracy over sampling speed by not selecting the fastest RQ-NSF implementation. Nonetheless, even the slowest model, \textbf{L}, achieves a sampling rate of $10$ million samples in under $12$ minutes on an NVIDIA RTX-4090 GPU and in $1$ hour $52$ minutes on a single Intel Xeon CPU core, far surpassing the requirement of $10$ million samples per day. This encourages us to consider this implementation as a viable approach for higher-dimensional processes, such as $2p2h$ or charged-current single-pion production.

\begin{table}[h!]
\centering

\begin{tabular}{c|ccc}
\toprule
\textbf{Model}                 & \textbf{S}   & \textbf{M}   & \textbf{L}    \\
 & & &\\ \midrule
Size (MB)                      & 200          & 719          & 1797          \\
 Number of flows& 10& 10&25\\
 Number of hidden nodes& 256& 512&512\\ 
Training time (hour)                      & 5          & 6          & 13          \\ 
CPU Speed (sec / million samples)  & 141.8         & 235.6         & 669.3\\ 
GPU Speed (sec / million samples)  & 12.5         & 23.6         & 70.7          \\ 
RESS, Flat Flux (\%)           & 98.48        & 98.64        & 98.87         \\ 
RESS, T2K Flux (\%)            & 98.48        & 98.64        & 98.85         \\ 
$\text{RMS}_z$ ($1s_{\sfrac{1}{2}}$)& 1.36& 1.01& 0.75\\ 
$\text{RMS}_z$ $(1p_{\sfrac{3}{2}}$)& 1.86& 1.52& 0.88\\ 
\bottomrule
\end{tabular}
\caption{Comparison of model performance for different model sizes (\textbf{S}, \textbf{M}, \textbf{L}). The size is here given for the eight models (as discussed in Section \ref{detail of implementation}).}
\label{tab:model_comparison}
\end{table}

The results show that models with larger sizes provide higher accuracy, as indicated by a steady rise in RESS values from 98.48\% (\textbf{S}) to 98.87\% (\textbf{L}). However, this improvement is relatively small in terms of unweighting efficiency in the case of important sampling, suggesting that the smallest model might be sufficient when sampling speed or memory constraints outweigh minor accuracy gains in the proposal distribution. This trade-off between sampling speed and accuracy depends on how the surrogate cross section is utilized. 

There are three ways to utilize the surrogate cross section:
\begin{enumerate}
    \item \emph{Use the surrogate as a proposal distribution} and apply an accept-reject algorithm to resample its outputs.
    \item \emph{Sample from the surrogate and provide per-event weights} by calling the true cross section once for each sampled event.
    \item \emph{Sample directly from the surrogate without any further resampling}. This option is by far the fastest for cross sections that are computationally heavy. However, it is important to mention that no generative model can replace a true cross section without introducing bias. Therefore, for a strictly unbiased analysis, one should always rely on the importance weight. As mentioned in Section~\ref{Comparison Density Estimation}, this is particularly crucial for studies focused on regions of very low cross section (e.g., low proton kinetic energy or backward-going protons with respect to the momentum transfer) or when very fine resolution is needed, such as angular resolutions better than one degree or energy resolutions finer than $1$~MeV.
\end{enumerate}

 Nevertheless, using the surrogate cross section for accept-reject sampling can be beneficial in some situations. In cases where the true cross section can be evaluated quickly, it may be most efficient to adopt the smallest model (\textbf{S}) and, if necessary, reweight its samples with an accept-reject algorithm. However, if evaluating the true cross section is time-consuming, the larger (\textbf{L}) model’s higher accuracy can justify its additional computational and storage costs. 
 
 Moreover, our normalizing flows approach to build a proposal distribution already achieves greater accuracy than the previous normalizing flows applied to neutrino-nucleus cross section modeling: Pina-Otey \emph{et al.}~\cite{PhysRevD.102.013003} reported a RESS of  $91.40\%$, which is significantly lower than even our smallest model (\textbf{S}). The concentration of weights around $1$ as shown in  Figure \ref{fig:weights} highlights the efficiency of the surrogate cross section as a proposal distribution for an accept-reject algorithm. The low spread of weights ensures that most samples contribute effectively to the estimation of observables, minimizing the variance introduced by the importance weights. If the weight threshold is set at $1.90$, corresponding to the $99.99^\text{th}$ percentile of weights then only one in ten thousand samples surpasses this limit. The expected acceptance rate for the surrogate cross section with this weight cap, when used as a proposal distribution, is approximately $ \frac{1}{1.90} \approx 53\% $.

Although the surrogate can serve as an efficient proposal distribution, Section~\ref{Comparison Datasets} shows that a dataset of events sampled directly from the normalizing flow surrogate is practically equivalent to an unbiased accept-reject dataset at our target precision. Section~\ref{Comparison Density Estimation} and Appendix~\ref{Profile appendix} further show that, once experimental conditions are taken into account, particularly a reasonable resolution in observables, the resulting uncertainties are already at the percent level for a monoenergetic neutrino beam. This residual error would be further reduced after convolution with a realistic neutrino flux, due to the additional smearing over neutrino energies. Importantly, the remaining modeling uncertainty is already well below the typical systematic uncertainties associated with experimental measurements, making the Normalizing Flows surrogate sufficiently accurate for practical analyses. This is particularly advantageous for complex, theory-driven cross sections requiring lengthy evaluations, since normalizing flows can reduce total sampling time from several days on a CPU (via accept-reject) to a few minutes on a GPU or CPU.

\section{Discussion}
\label{discussion}

The aim of this study was to demonstrate the ability to efficiently and accurately generate $1p1h$ samples without relying on the full complexity of the Ghent model, which also accounts for two-body currents and the Continuum Random Phase Approximation (CRPA)~\cite{PhysRevC.65.025501}. These corrections are not expected to pose major challenges for the proposed method. Two-body currents, involving two-particle states, provide corrections to the impulse approximation, correcting the single nucleon knockout processes and giving rise to double nucleon knockout processes. The effect of two-body currents on the $1p1h$ cross section is relatively small and can be accounted for by fine-tuning a model already trained within the impulse approximation.

Similarly, although including more detailed physics features in the cross section, from a numerical point of view the inclusion of CRPA correlations is expected to simplify the energy transfer distribution by broadening the sharp features at low energy, smoothing out the delta-like structures associated with the missing energy from shell modeling, hence facilitating event generation using normalizing flows. 

The greater challenge arises in describing $2p2h$ processes due to the increased number of dimensions in the phase space. 
These processes correspond to two-nucleon knockout, and their fully exclusive differential cross sections involve nine independent kinematic variables:
\begin{equation*}
\frac{d^9\sigma (E_\nu)}{d\omega \, d\Omega_\mu \, d\Omega_{N_1} \, dE_{N_1} \,  d\Omega_{N_2} \, dE_{N_2}} \propto L_{\mu\nu} W^{\mu\nu}.
\end{equation*}

As in the $1p1h$ case, we treat the neutrino energy as a conditional variable, the differential cross section does not depend on the muon's azimuthal angle $\phi$, and we incoherently sum over the energy eigenstates of the target nucleons. Consequently, we must handle seven independent kinematic variables, three more than in the $1p1h$ scenario. The increased dimensionality should not affect the expressiveness of autoregressive normalizing flows as shown in~\cite{Coccaro_2024}. However, a larger phase space requires more training samples, and this challenge is made even more difficult by the longer computation time needed for the $2p2h$ cross section.

In this context, our work on $1p1h$ modeling provides a solid foundation for extending the approach to semi-exclusive cross sections, where the kinematics of the subleading nucleon are integrated out. Training a model on semi-exclusive cross sections captures key features of the $2p2h$ process and allows for a more efficient training of fully exclusive $2p2h$ cross sections. Instead of using uniformly distributed samples, as done in this work, the training can be performed using samples drawn from the surrogate semi-exclusive cross section, reducing the number of required evaluations and improving training efficiency.

\section{Conclusion}

In this work, we presented a normalizing flow-based surrogate for generating neutrino-nucleus scattering events with high accuracy and efficiency. By comparing our surrogate’s output to that of an unbiased accept-reject method, we demonstrated that the two methods produce practically equivalent samples across a wide range of neutrino energies and nuclear shells. Even more importantly, we showed that the normalizing flow approach can generate millions of events in minutes on modern GPUs. 

From a theoretical standpoint, such surrogates enable systematic studies of model variations and uncertainty quantification, giving insights into how different theoretical inputs manifest across multi-dimensional phase space.

Looking ahead, the flexibility of normalizing flows makes them well-suited to other processes beyond the $1p1h$ process studied here. In particular, higher-dimensional processes such as $2p2h$ or single-pion production can benefit from a similar surrogate approach. Currently, neutrino experiments such as T2K rely on so-called "Frankenstein" models that are combinations of submodels drawn from different theoretical frameworks, each tuned to specific interaction channels and combined using ad-hoc relative proportions. While practical, this approach makes it difficult to disentangle whether observed features in predictions arise from genuine physical effects, differences in underlying frameworks, or purely phenomenological adjustments. In contrast, a normalizing flow-based surrogate can make a unified sampling scheme for \emph{all} processes within a single, theory-grounded cross section model possible. It could do so even in the case of a mean-field model, such as that developed by the Ghent group, which has been before too computationally intensive to integrate directly into event generators in its full complexity. Such a tool could be crucial for future neutrino experiments, delivering fast, accurate, and cohesive event generation that seamlessly bridges the gap between the experimental and theoretical frontiers.

\section*{Acknowledgments}
M.E. acknowledges support from the Swiss National Foundation grants No. $200021E\_213196$.
A.N. is supported by the Neutrino Theory Network under Award Number DEAC02-07CH11359, the department of physics and nuclear theory group at the University of Washington.
K.N. acknowledges support from the Fund for Scientific Research Flanders (FWO) and Ghent University Special Research Fund. M.E. finally thanks Joshua Isaacson for the valuable discussion they had during the end of the writing of this article.

\bibliography{apssamp}

\appendix
\section{Profile of the proton kinetic energy for different angles of the outgoing muon and proton}
\label{Profile appendix}

To complement the analysis presented in Section~\ref{density finite resolution}, this appendix illustrates the impact of the biases between the true and surrogate cross sections on the distribution of proton kinetic energy $T_N$ at fixed kinematics, both with and without local averaging. The study is performed using a monoenergetic neutrino beam with energy $E_\nu = 650\,\text{MeV}$, and the total cross section is computed by summing the contributions from the two occupied shells, as described in Equation~\ref{sum partial}. In the case of CCQE interaction in the impulse approximation, one can simply derive the kinetic energy by computing : 
$$T_N = \omega - E_b$$
where $E_b$ is the shell dependent bounding energy. 

Figures \ref{fig:TN5}, \ref{fig:TN20} and \ref{fig:TN50} show the comparisons between true and surrogate cross sections for $\theta_N = 5^\circ$, $20^\circ$, and $50^\circ$, respectively. In each figure, the top panel displays the differential cross section as a function of $T_N$ for various values of $\theta_\mu$, comparing the NF surrogate to the true differential cross section. The corresponding relative difference $\Delta$ is also shown. In the bottom panel, the same comparison is shown after averaging the cross section over a local 4D kinematic cube centered at the same point in $(T_N, \theta_\mu, \theta_N, \phi_N)$ space, using a bin size of $1$~MeV in $T_N$ and $1^\circ$ in each angular variable. 

The top panels reveal small but noticeable differences between the true cross section and the prediction from the Normalizing Flows, both in the tail and in the core of the distribution. However, most of these differences, particularly in the core, are significantly reduced when averaging over a small neighborhood in kinematic space as illustrated in the bottom plots. This supports the conclusion that the remaining differences primarily reflect localized fluctuations rather than global modeling errors. Smaller deviations persist after averaging particularly in regions of low cross section. For analyses that probe regions of very low cross section (e.g., backward-going protons with respect to the momentum transfer or low $T_N$) or require sub-degree angular precision or sub-MeV energy resolution, the surrogate model may introduce non-negligible bias, making it necessary to reweight the events using the true cross section. 

At the same time, the surrogate retains distinctive kinematic patterns predicted by the Ghent model, inherited from the shell modeling and the distorted wave approach. Using a generator whose truth distribution is matched to that surrogate model will therefore show these signatures in simulated samples, increasing the chances that experiments with near-$4\pi$ angular acceptance and high‑granularity detectors, with good sensitivity to low-momentum protons, such as the SuperFGD at T2K~\cite{Blondel_2020}, will be able to observe some of them, if they are not already visible in existing datasets.

\begin{figure*}
\includegraphics[width=0.99\linewidth, trim=0cm 0cm 0cm 0cm, clip]{Images/1p1h_thetaN_5_TN.png}
\caption{
Profile of the proton kinetic energy distribution $T_N$ for fixed values of the muon scattering angle $\theta_\mu \in \{10^\circ, 20^\circ, 50^\circ, 80^\circ, 120^\circ, 170^\circ\}$, and for fixed outgoing nucleon angles $(\theta_N, \phi_N) = (5^\circ, 180^\circ)$. The cross section is computed for a monoenergetic neutrino beam of $650$~MeV and for both shells combined. The top panel displays the differential cross section as a function of $T_N$ for each $\theta_\mu$ value, comparing the output of the NF surrogate to the true model. The NF prediction is shown with dashed lines, and the true model with solid lines. The relative difference $\Delta$ between the two curves is also shown. In the bottom panel, the same comparison is shown after averaging the cross section over a local 4D kinematic cube centered at the same $(T_N, \theta_\mu, \theta_N, \phi_N)$ point. The bin resolution of this cube is $1$~MeV in $T_N$ and $1^\circ$ in each angular dimension. Each cube contains $500$ uniformly distributed events sampled from the NF or from the true cross section.
}

\label{fig:TN5}
\end{figure*}

\begin{figure*}[t]
  \centering
  \includegraphics[width=\linewidth]{Images/1p1h_thetaN_20_TN.png}
  \caption{ $\theta_N = 20^\circ$}
  \label{fig:TN20}
\end{figure*}

\begin{figure*}[t]
  \centering
  \includegraphics[width=\linewidth]{Images/1p1h_thetaN_50_TN.png}
  \caption{%
     $\theta_N = 50^\circ$.
  }
  \label{fig:TN50}
\end{figure*}

\end{document}